\documentclass[letterpaper,10pt,conference]{ieeeconf}

\IEEEoverridecommandlockouts
\usepackage{amsmath,amssymb,amsfonts}
\usepackage{mathtools}
\usepackage{bm}
\usepackage{graphicx}
\usepackage{booktabs}
\usepackage{cite}
\usepackage{xcolor}
\usepackage{balance}
\newtheorem{theorem}{Theorem}
\newtheorem{proposition}{Proposition}
\newtheorem{corollary}{Corollary}
\newtheorem{assumption}{Assumption}

\DeclareMathOperator*{\argmax}{arg\,max}

\title{\LARGE \bf
Shared-Harmonic Characterization and Robust Orientation-Selection Certification for Directional Sensing Under Anisotropic Target Motion}

\author{Mingyu Kim$^{*}$, Daniel J. Stilwell, and Jorge G. Jimenez%
\thanks{M. Kim is with the Department of Electrical and Computer Engineering,
Georgia Southern University, Statesboro, GA 30460, USA
(e-mail: mkim@georgiasouthern.edu).}%
\thanks{D. J. Stilwell is with the Bradley Department of Electrical and
Computer Engineering, Virginia Tech, Blacksburg, VA 24061, USA.}%
\thanks{J. G. Jimenez is with The Johns Hopkins University Applied Physics
Laboratory, Laurel, MD 20723, USA.}%
\thanks{$^{*}$Corresponding author.}%
}

\begin{document}

\maketitle
\thispagestyle{empty}
\pagestyle{empty}

\begin{abstract}
Directional sensors are often oriented using historical target trajectories, while future trajectories remain uncertain at deployment. This letter addresses two post-placement questions: i) when orientation materially affects detection and ii) when a historically selected orientation remains reliable under future target-motion shift. We develop a placement-aware framework for anisotropy-induced detection (AID), showing that orientation sensitivity is governed by harmonic modes shared by the locally visible target-direction distribution and the sensor trajectory-incidence response. The resulting AID measure admits an exact RMS interpretation for the retained-harmonic orientation response. We then introduce a sensor-induced directional discrepancy, derive a conditional bound on future orientation regret, and obtain sufficient certificates for robust orientation selection whenever the future sensor-relevant shift lies within a prescribed discrepancy radius. Using 2025 AIS ship-traffic data near Terrebonne Bay, Louisiana, historical AID is strongly correlated with held-out orientation sensitivity ($r=0.908$).
Across four angular-response widths, AID explains
orientation-sensitivity variation with $R^2=0.920$,
compared with $R^2=0.239$ for target-only anisotropy.
At the nominal threshold, 316 of 350 locations are empirically classified as orientation-useful, and all retain gain
above the threshold across five held-out windows.

\end{abstract}

\section{Introduction}
\label{sec:intro}

Directional and aspect-sensitive sensing performance depends
on sensor orientation relative to target motion or incidence angle.
Directional-sensor coverage, sensor selection, and orientation
optimization have been studied extensively
\cite{guvensan2011survey,ai2006coverage,fusco2009selection}.
Related approaches jointly optimize position and orientation or incorporate
probabilistic directional sensing models
\cite{akbarzadeh2014gradient,an2021anisotropic,ju2022directional,
si2019probabilistic}. These formulations primarily address
\emph{how a sensor should be oriented} once a target or coverage model is
specified.

Target motion introduces an additional source of uncertainty. Existing
methods incorporate probabilistic mobile-target models or adapt sensing
directions as targets evolve \cite{wasim2020mobile,son2026mobile}, while
informative-sensing and stochastic-placement methods optimize sensor
locations or trajectories under uncertain target occurrence and motion
\cite{krause2006near,hollinger2014sampling}. In our prior work, Poisson
target models were used for spatial sensor placement \cite{kim2023poisson},
and stochastic target trajectories were subsequently considered for
near-optimal barrier placement using historical maritime traffic
\cite{kim2025barrier}. These studies emphasize placement or
orientation optimization rather than whether orientation selection
remains worthwhile and reliable after placement is fixed.

Orientation sensitivity depends jointly on placement, target
motion, and sensor response. Placement filters the trajectory field
into a locally visible directional density, whose alignment with
the sensor's angular response determines orientation-dependent
detection. Thus, comparable target exposure or strong target
anisotropy alone does not imply comparable orientation sensitivity:
the target distribution and sensor response must share relevant
directional structure. We term this coupled variation
\emph{anisotropy-induced detection (AID)}.

This letter addresses two post-placement questions:
i) whether orientation materially affects detection and
ii) whether an orientation selected from historical trajectories
remains beneficial under future target-motion shift.
The objective is orientation selection, not dynamic control
synthesis; orientation dynamics and feedback are not modeled.
We characterize the orientation response through circular
correlation and shared Fourier harmonics, then use a sensor-aware
distribution-shift measure to assess the robustness of historical
orientation decisions.

\subsubsection*{Contributions}
The contributions are threefold.
First, we establish a necessary and sufficient shared-harmonic
condition for orientation-dependent detection, explicitly
coupling placement-visible target anisotropy with the sensor
response. The contribution is its sensing-specific
interpretation, not the underlying Fourier identity.
Second, we define a spectral AID measure and establish its
exact relationship to the RMS orientation-response variation.
Third, we introduce a sensor-induced directional discrepancy
and derive conditional bounds on future orientation regret
and sufficient conditions for retaining a historically selected
orientation. Data-calibrated conditions are evaluated
empirically on strictly held-out trajectories rather than
treated as unconditional future guarantees.

\section{Placement-Aware Directional Detection}
\label{sec:model}

Consider a bounded surveillance region
$\Omega\subset\mathbb{R}^{2}$ and represent each locally linear
target-trajectory segment by an oriented line
\begin{equation*}
\ell(\xi,p)
=
\left\{
p\mathbf n(\xi)+s\mathbf t(\xi):s\in\mathbb{R}
\right\},
\label{eq:oriented_line}
\end{equation*}
where
\[
\mathbf t(\xi)=
\begin{bmatrix}\cos\xi&\sin\xi\end{bmatrix}^{\top},
\qquad
\mathbf n(\xi)=
\begin{bmatrix}-\sin\xi&\cos\xi\end{bmatrix}^{\top}.
\]
Here, $\xi\in[0,2\pi)$ denotes the direction of travel and
$p$ is the signed perpendicular offset from a fixed origin.
The trajectory space is
$\Psi=[0,2\pi)\times\mathcal P$, where
$\mathcal P=[-p_{\max},p_{\max}]$ covers trajectories intersecting
$\Omega$, which are defined in \cite{kim2026sensors}. Nonlinear trajectories may be partitioned into short segments and
represented in the same space, as in \cite{kim2025barrier}.


Let $\lambda(\xi,p)\geq0$ denote the trajectory intensity over $\Psi$, so
that $\lambda(\xi,p)\,d\xi\,dp$ represents the expected trajectory occurrence
in direction--offset space. 
The theoretical results apply to any trajectory
intensity model satisfying the stated regularity
conditions, without requiring a specific estimation method.


Sensor $i$ is located at $a_i\in\Omega$ and has controllable orientation
$\phi_i\in[0,2\pi)$.
Its placement-dependent trajectory availability
$\gamma_i(\xi,p;a_i)\in[0,1]$ accounts for spatial effects
such as range and occlusion. Importantly, $\gamma_i$ may
depend on target direction $\xi$ and offset $p$, but not
on controllable sensor orientation $\phi_i$.

Orientation-dependent detection quality is modeled by
$h_i(\xi-\phi_i;\chi_i)\in[0,1]$, where $h_i$ is measurable
and $2\pi$-periodic, and $\chi_i$ specifies angular acceptance.
Angular differences are interpreted modulo $2\pi$.
The model isolates direction-dependent sensing quality
under fixed spatial visibility, motivated by aspect-sensitive
radar or sonar sensing \cite{marques2012directional,choo2021bistatic}.
Relative travel direction is a modeled angular variable
and need not equal physical target aspect.
Sensors whose spatial visibility changes with orientation,
such as rotating narrow-FOV cameras, fall outside this
separable formulation and require
$\gamma_i(\xi,p;a_i,\phi_i)$.

We assume the separable trajectory-level detection model
\begin{equation*}
q_i(\xi,p;a_i,\phi_i,\chi_i)
=
\gamma_i(\xi,p;a_i)
h_i\!\left(\xi-\phi_i;\chi_i\right).
\label{eq:q}
\end{equation*}

This separability distinguishes placement-dependent
trajectory observability from orientation-dependent
sensing quality and underlies the shared-harmonic analysis.

Conditioned on $\lambda$, the expected detected trajectory intensity is
\begin{equation}
\begin{aligned}
D_i(a_i,\phi_i;\chi_i)
&=
\int_{\mathcal P}\int_{0}^{2\pi}
\lambda(\xi,p)\gamma_i(\xi,p;a_i) \\
&\qquad\cdot
h_i\!\left(
\xi-\phi_i;\chi_i
\right)
\,d\xi\,dp .
\end{aligned}
\label{eq:D}
\end{equation}

Define the \emph{placement-visible trajectory intensity}
\begin{equation*}
Z_i(a_i)
=
\int_{\mathcal P}\int_{0}^{2\pi}
\lambda(\xi,p)\gamma_i(\xi,p;a_i)
\,d\xi\,dp ,
\label{eq:Z}
\end{equation*}
which is independent of $\phi_i$.
We restrict attention to $0<Z_i(a_i)<\infty$;
otherwise the normalized orientation problem is undefined.

For $Z_i(a_i)>0$, define the normalized detection performance
\begin{equation}
d_i(a_i,\phi_i;\chi_i)
=
\frac{D_i(a_i,\phi_i;\chi_i)}
     {Z_i(a_i)}.
\label{eq:d}
\end{equation}
Since $0\leq h_i\leq1$, we have $0\leq d_i\leq1$;
normalization removes differences in visible traffic intensity.

The corresponding \emph{locally visible directional density} is
\begin{equation}
f_i(\xi;a_i)
=
\frac{
\displaystyle
\int_{\mathcal P}
\lambda(\xi,p)\gamma_i(\xi,p;a_i)\,dp
}{
Z_i(a_i)
},
\qquad
Z_i(a_i)>0,
\label{eq:f}
\end{equation}
with $\int_0^{2\pi}f_i(\xi;a_i)\,d\xi=1$.
Substitution into \eqref{eq:d} yields
\begin{equation}
d_i(a_i,\phi_i;\chi_i)
=
\int_{0}^{2\pi}
f_i(\xi;a_i)
h_i\!\left(
\xi-\phi_i;\chi_i
\right)
\,d\xi .
\label{eq:correlation}
\end{equation}
Equation~\eqref{eq:correlation} expresses normalized detection
as a circular correlation: placement determines $f_i$,
while orientation shifts $h_i$. Consequently, comparable
visible trajectory intensities $Z_i$ need not imply comparable
orientation sensitivity, which Section~\ref{sec:aid}
characterizes through shared harmonics.

\section{When Does Orientation Matter?}
\label{sec:aid}

We characterize when sensor orientation affects
normalized detection and quantify this dependence
through shared target--sensor Fourier harmonics.
Fix sensor location $a_i$ and angular-acceptance parameter
$\chi_i$, suppressing them for brevity. From
\eqref{eq:correlation},
\begin{equation}
d_i(\phi)
=
\int_{0}^{2\pi}
f_i(\xi)h_i(\xi-\phi)\,d\xi .
\label{eq:conditional_correlation}
\end{equation}
Since $h_i$ is measurable and bounded,
$h_i\in L^2([0,2\pi))$. For $f_i$, we require
$f_i\in L^2([0,2\pi))$. A sufficient condition is
$\lambda\gamma_i\in L^\infty(\Psi)$ and
$0<Z_i(a_i)<\infty$: if $\lambda\gamma_i\leq M_i$
almost everywhere, then
$f_i(\xi;a_i)\leq
|\mathcal P|M_i/Z_i(a_i)$ almost everywhere,
which implies $f_i\in L^2([0,2\pi))$.
Using standard real Fourier series
\cite{stein2003fourier},
\begin{align}
f_i(\xi)
&=
\frac{1}{2\pi}
+
\sum_{m=1}^{\infty}
F_{i,m}
\cos\!\left(m\xi-\theta^f_{i,m}\right),
\label{eq:f_fourier}\\
h_i(\nu)
&=
c_{i,0}
+
\sum_{m=1}^{\infty}
H_{i,m}
\cos\!\left(m\nu-\theta^h_{i,m}\right),
\label{eq:h_fourier}
\end{align}
where
$c_{i,0}=(2\pi)^{-1}\int_0^{2\pi}h_i(\nu)\,d\nu$.
The Fourier magnitudes and phases are
$F_{i,m}=\sqrt{\alpha_{i,m}^2+\beta_{i,m}^2}$,
$\theta^f_{i,m}=\operatorname{atan2}(\beta_{i,m},\alpha_{i,m})$,
with
\[
\alpha_{i,m}=\frac{1}{\pi}
\int_0^{2\pi}f_i(\xi)\cos(m\xi)\,d\xi,
\quad
\beta_{i,m}=\frac{1}{\pi}
\int_0^{2\pi}f_i(\xi)\sin(m\xi)\,d\xi.
\]
Analogously, $H_{i,m}$ and $\theta^h_{i,m}$ are
the magnitude and phase of the corresponding Fourier
coefficients of $h_i$. Phases of zero-magnitude modes
are arbitrary.

\begin{theorem}[Shared-harmonic characterization]
\label{thm:shared}
For a fixed sensor location with $Z_i(a_i)>0$,
\begin{equation}
d_i(\phi)
=
c_{i,0}
+
\pi
\sum_{m=1}^{\infty}
F_{i,m}H_{i,m}
\cos\!\left(
m\phi+\theta^h_{i,m}-\theta^f_{i,m}
\right).
\label{eq:d_harmonic}
\end{equation}
Consequently, $d_i(\phi)$ is independent of orientation if and only if
\begin{equation}
F_{i,m}H_{i,m}=0,
\qquad
\forall m\geq1.
\label{eq:shared_condition}
\end{equation}
\end{theorem}

\begin{proof}
The constant contribution is $c_{i,0}$ because
periodicity gives
$\int_0^{2\pi}h_i(\xi-\phi)\,d\xi=2\pi c_{i,0}$
for every $\phi$.
Substituting \eqref{eq:f_fourier}--\eqref{eq:h_fourier}
into \eqref{eq:conditional_correlation}, Fourier
orthogonality eliminates unequal harmonic orders,
while matching order $m$ contributes
$\pi F_{i,m}H_{i,m}
\cos(m\phi+\theta^h_{i,m}-\theta^f_{i,m})$.
This gives \eqref{eq:d_harmonic}.
Uniqueness of Fourier coefficients then implies that
$d_i(\phi)$ is constant if and only if every nonzero-order
coefficient vanishes, yielding \eqref{eq:shared_condition}.
\end{proof}

Thus, orientation dependence requires shared nonzero
harmonics: placement determines $F_{i,m}$ through $f_i$,
whereas the sensor response determines $H_{i,m}$.
Averaging \eqref{eq:d_harmonic} over orientation gives
\begin{equation}
\bar d_i
:=
\frac{1}{2\pi}
\int_0^{2\pi}d_i(\phi)\,d\phi
=
c_{i,0},
\label{eq:mean_detection}
\end{equation}
which provides an orientation-independent
normalized-detection baseline. Thus, $c_{i,0}$ is the expected normalized
detection performance under a uniformly random
sensor orientation, providing an orientation-neutral
baseline rather than a fixed-orientation or
dynamic-control benchmark.

Define the shared-mode strength
$e_{i,m}:=F_{i,m}^2H_{i,m}^2$.
For retained harmonic order $K$, the
anisotropy-induced detection (AID) level is
\begin{equation}
\rho_{\mathrm{AID},i}^{(K)}
:=
\left(
\sum_{m=1}^{K}e_{i,m}
\right)^{1/2}.
\label{eq:rho}
\end{equation}

\begin{proposition}[Exact RMS interpretation of AID]
\label{prop:rms}
Let $d_i^{(K)}(\phi)$ denote
\eqref{eq:d_harmonic} truncated at order $K$. Then
\begin{equation}
\left[
\frac{1}{2\pi}
\int_{0}^{2\pi}
\left(
d_i^{(K)}(\phi)-c_{i,0}
\right)^2
d\phi
\right]^{1/2}
=
\frac{\pi}{\sqrt{2}}\,
\rho_{\mathrm{AID},i}^{(K)}.
\label{eq:rms}
\end{equation}
\end{proposition}

\begin{proof}
Parseval's identity gives
\[
\frac{1}{2\pi}\int_0^{2\pi}
\left(d_i^{(K)}(\phi)-c_{i,0}\right)^2d\phi
=
\frac{\pi^2}{2}
\sum_{m=1}^{K}F_{i,m}^2H_{i,m}^2,
\]
which yields \eqref{eq:rms}.
\end{proof}

Theorem~\ref{thm:shared} characterizes whether
orientation affects detection, while
Proposition~\ref{prop:rms} quantifies its RMS variation
through the coupled target--sensor spectrum.

For numerical implementation, let
$E_{i,\max}:=\sum_{m=1}^{K_{\max}}e_{i,m}$, where
$K_{\max}$ is the largest resolved harmonic order.
For $E_{i,\max}>0$, select
\begin{equation}
K_i^\star
=
\min
\left\{
K:
\frac{\sum_{m=1}^{K}e_{i,m}}
     {E_{i,\max}}
\geq\eta
\right\},
\qquad
\eta\in(0,1].
\label{eq:Kstar}
\end{equation}
If $E_{i,\max}=0$, set $K_i^\star=0$.
Thus, $K_i^\star$ retains at least an $\eta$ fraction
of the resolved shared-mode strength.

\section{Sensor-Aware Historical-to-Future Robustness}
\label{sec:robust}
In this section, we assess whether orientations selected from historical
trajectories remain effective under future changes in
the locally visible target-direction distribution.
Let $\hat f_i$ denote the locally visible directional density estimated from
historical trajectories and let $f_i^+$ denote the corresponding future
density. Their orientation-dependent performances are
\begin{align*}
\hat d_i(\phi)
&=
\int_0^{2\pi}
\hat f_i(\xi)h_i(\xi-\phi)\,d\xi,\\
d_i^+(\phi)
&=
\int_0^{2\pi}
f_i^+(\xi)h_i(\xi-\phi)\,d\xi.
\end{align*}

To measure sensing-relevant distribution shift,
define the sensor-induced directional discrepancy
for arbitrary directional densities $j$ and $g$:
\[
\mathcal D_i(j,g)
:=
\sup_{\phi\in[0,2\pi)}
\left|
\int_0^{2\pi}
[j(\xi)-g(\xi)]h_i(\xi-\phi)\,d\xi
\right|.
\]

This is an IPM-type discrepancy over the rotated
sensor responses $\{h_i(\cdot-\phi):\phi\in[0,2\pi)\}$
\cite{muller1997ipm}. In particular,
\begin{equation}
\mathcal D_i(f_i^+,\hat f_i)
=
\sup_{\phi}
\left|
d_i^+(\phi)-\hat d_i(\phi)
\right|.
\label{eq:Dh_equiv}
\end{equation}

\begin{assumption}[Bounded sensor-aware shift]
\label{ass:Dh}
For some $\epsilon_i\geq0$, the future distribution
satisfies
\[
\mathcal D_i(f_i^+,\hat f_i)\leq\epsilon_i.
\]
\end{assumption}
By \eqref{eq:Dh_equiv}, Assumption~\ref{ass:Dh}
implies the uniform orientation-response bound
\begin{equation}
\left|
d_i^+(\phi)-\hat d_i(\phi)
\right|
\leq
\epsilon_i,
\qquad
\forall\phi\in[0,2\pi).
\label{eq:uniform_bound}
\end{equation}

\begin{theorem}[Future orientation-regret bound]
\label{thm:robust}
Let
$\hat\phi_i^\star\in\argmax_\phi\hat d_i(\phi)$
and
$\phi_i^{+,\star}\in\argmax_\phi d_i^+(\phi)$.
Under Assumption~\ref{ass:Dh},
\[
0
\leq
d_i^+(\phi_i^{+,\star})
-
d_i^+(\hat\phi_i^\star)
\leq
2\epsilon_i.
\]
\end{theorem}

\begin{proof}
Using \eqref{eq:uniform_bound},
\[
\begin{aligned}
d_i^+(\phi_i^{+,\star})
-
d_i^+(\hat\phi_i^\star)
&\leq
\hat d_i(\phi_i^{+,\star})
-
\hat d_i(\hat\phi_i^\star)
+
2\epsilon_i \\
&\leq
2\epsilon_i,
\end{aligned}
\]
because $\hat\phi_i^\star$ maximizes $\hat d_i$. Nonnegativity follows from
the optimality of $\phi_i^{+,\star}$ for $d_i^+$.
\end{proof}

Thus, the historical orientation need not remain
future-optimal to have bounded regret.
To determine whether retaining it is worthwhile,
define gains relative to the orientation-average baseline
$c_{i,0}$ in \eqref{eq:mean_detection}:
\[
\hat G_i
:=
\max_{\phi}\hat d_i(\phi)-c_{i,0},
\qquad
G_i^+
:=
\max_{\phi}d_i^+(\phi)-c_{i,0}.
\]
The baseline is common to both distributions
because $h_i$ is unchanged.

\begin{corollary}[Orientation-selection certificate]
\label{cor:certificate}
Under Assumption~\ref{ass:Dh},
\[
|G_i^+-\hat G_i|
\leq
\epsilon_i.
\]
Furthermore, the historical orientation satisfies
\begin{equation}
d_i^+(\hat\phi_i^\star)-c_{i,0}
\geq
\hat G_i-\epsilon_i.
\label{eq:historical_orientation_gain}
\end{equation}
For a minimum useful orientation gain $\tau>0$,
\begin{align}
\hat G_i+\epsilon_i<\tau
&\Rightarrow
G_i^+<\tau,
\label{eq:no_control}\\
\hat G_i-\epsilon_i>\tau
&\Rightarrow
d_i^+(\hat\phi_i^\star)-c_{i,0}>\tau.
\label{eq:active_control}
\end{align}
The first condition certifies that orientation selection
cannot achieve the threshold even with future-optimal
orientation; the second certifies that the historical
orientation exceeds it. Otherwise, the decision is
uncertain under the assumed discrepancy bound.
\end{corollary}

The deployment bound follows from
$d_i^+(\hat\phi_i^\star)
\geq\hat d_i(\hat\phi_i^\star)-\epsilon_i
=c_{i,0}+\hat G_i-\epsilon_i$.

In practice, estimate the unknown discrepancy radius
from calibration-window directional densities
$\tilde f_{i,w}$, $w\in\mathcal W$, using
\begin{equation}
\hat\epsilon_i
=
\max_{w\in\mathcal W}
\mathcal D_i(\tilde f_{i,w},\hat f_i).
\label{eq:eps_cal}
\end{equation}
This empirical radius does not guarantee that future
discrepancies remain below $\hat\epsilon_i$.
Thus, decisions using $\hat\epsilon_i$ are empirically
calibrated, not unconditional certificates;
Section~\ref{sec:numerics} evaluates their transfer
to strictly held-out trajectories.


\begin{figure}[t]
    \centering
    \includegraphics[width=0.5\columnwidth]{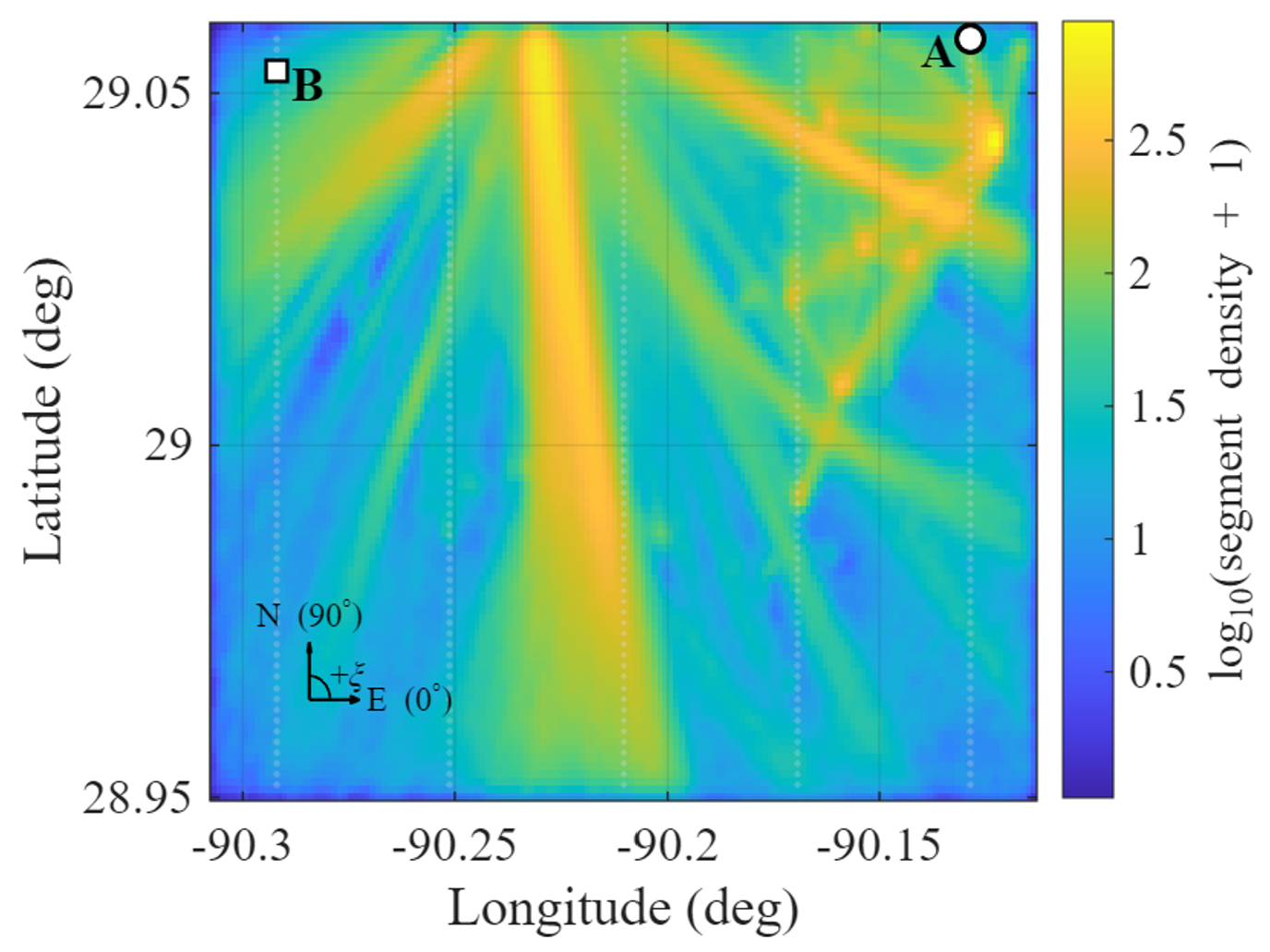}
    \caption{Historical reconstructed trajectory-segment density
    with the $70\times5$ candidate-location grid and matched
    locations A and B. Travel directions are measured
    counterclockwise from east, with $0^\circ$ corresponding
    to east and $90^\circ$ to north.}
    \label{fig:study_area}
\end{figure}

\section{AIS-Based Numerical Validation}
\label{sec:numerics}

In this section, we evaluate the proposed framework using 2025 Automatic Identification System (AIS) ship-traffic data near Terrebonne Bay, Louisiana. The dataset contains $1{,}314{,}015$ reports over $28.9495^\circ$--$29.0600^\circ$ N and $90.3077^\circ$--$90.1129^\circ$ W. Reports are grouped by Maritime Mobile Service Identity (MMSI), ordered in time, split when consecutive reports are separated by more than 30~min or 10~km, and resampled every 250~m. Passes shorter than 500~m are discarded. To reduce dependence on vessel-specific AIS reporting rates, each reconstructed pass contributes one locally weighted directional observation at its closest approach to each candidate sensor.

We evaluate 350 candidate locations on a
$70\times5$ uniform grid spanning the inner $96\%$
of the historical latitude range and $84\%$ of the
longitude range (Fig.~\ref{fig:study_area}).
This grid samples orientation sensitivity without
optimizing placement. Directions are measured
counterclockwise from east ($0^\circ$ east,
$90^\circ$ north).
 A 3-km visibility radius is used with Gaussian range weighting of standard deviation 1.5~km. Unless otherwise stated, the trajectory-incidence response is modeled using a tapered $60^\circ$ total angular-acceptance width,
\begin{equation}
h_i(\nu)
=
\begin{cases}
\cos^2\!\left(\dfrac{\pi\nu}{2\chi_i}\right),
& |\nu|\leq\chi_i,\\[1mm]
0, & \text{otherwise},
\end{cases}
\qquad
\chi_i=30^\circ .
\label{eq:numerical_sensor}
\end{equation}
The tapered $\cos^2$ profile serves as a smooth compact-support angular-acceptance model rather than a calibration of a particular sensing device. Direction is discretized into 72 bins and sensor orientation into 360 angles. We use $K_{\max}=18$, energy-retention level $\eta=0.99$, and nominal minimum useful normalized gain $\tau=0.05$.

The data are divided chronologically into Jan.--Aug. for historical estimation (901,046 reports), Sep.--Oct. for calibration (197,639 reports), and Nov.--Dec. for strictly held-out evaluation (215,330 reports). The historical period contains 25,945 reconstructed vessel passes. The calibration and held-out periods are each divided into five approximately equal 12--13 day windows so that the empirical discrepancy radius and future discrepancies are evaluated at matched temporal scales.

\subsection{Placement, Shared Harmonics, and AID}

\begin{figure}[t]
\vspace*{0.05in}
\centering
\includegraphics[width=0.47\textwidth]{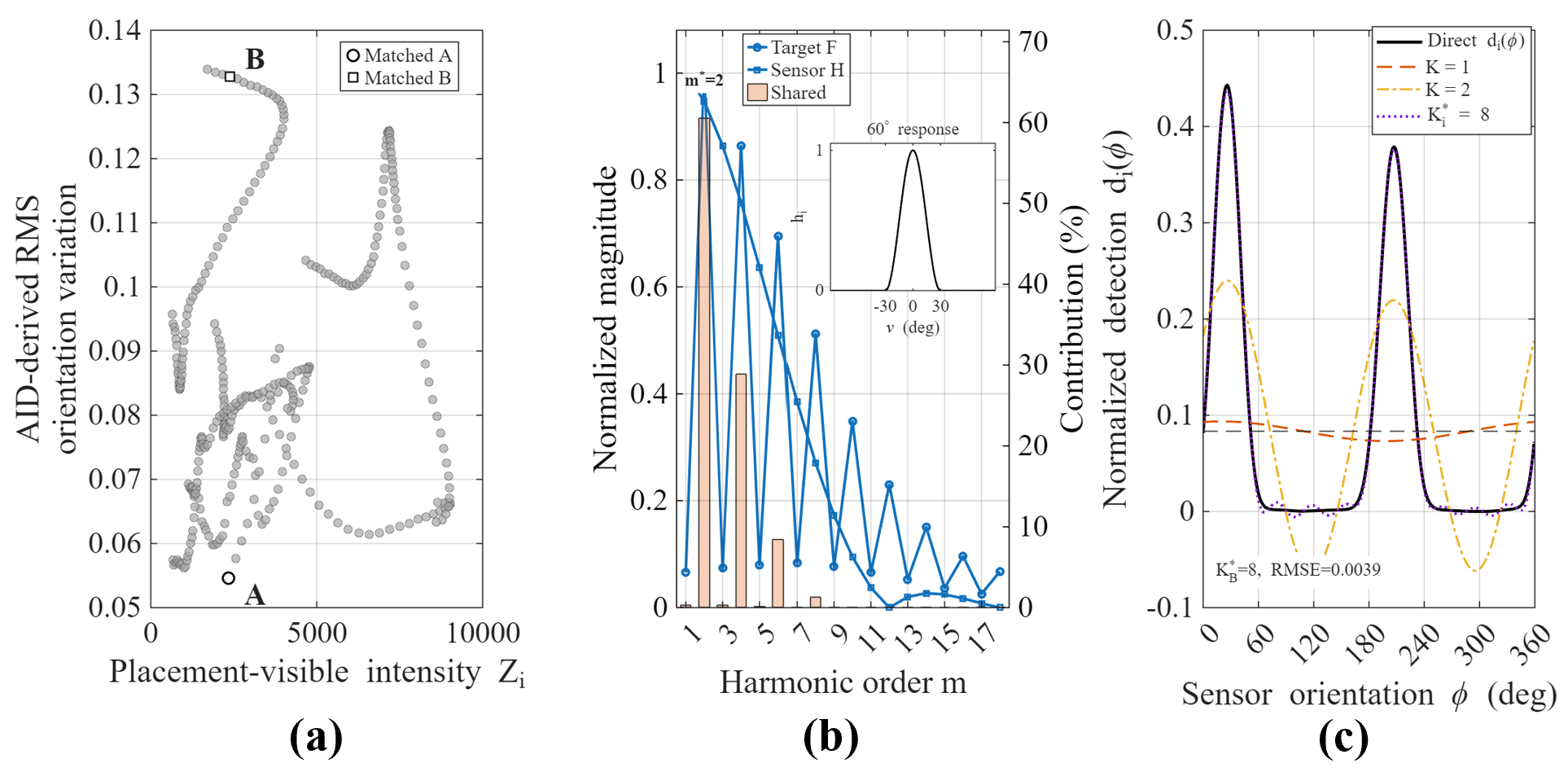}
\caption{Placement and shared-harmonic mechanisms underlying
AID. (a) AID-derived RMS orientation variation versus
placement-visible trajectory intensity for 350 candidate
locations (gray dots), with matched locations A and B
highlighted.
(b) Target and sensor harmonic magnitudes
and shared-mode contributions at location B;
the inset shows the nominal $60^\circ$ tapered angular response.
(c) Direct and truncated harmonic reconstructions of the
orientation response for $K=1$, $K=2$, and $K_B^\star=8$.}
\label{fig:aid_mechanism}
\end{figure}

Figure~\ref{fig:aid_mechanism}(a) plots
AID-derived RMS orientation variation against
placement-visible trajectory intensity for all
350 candidate locations (gray dots). It illustrates
that visible traffic intensity alone does not
determine orientation sensitivity, consistent
with Theorem~\ref{thm:shared} and
Proposition~\ref{prop:rms}.
Candidate pairs are restricted to at most $3\%$ symmetric relative mismatch in $Z_i$ and $10\%$ mismatch in local pass count. Among 1,028 qualifying pairs, locations A and B denote the lower- and higher-sensitivity members of the pair with the largest AID-derived RMS difference.

The selected locations have $Z_A=2340.69$ and $Z_B=2381.10$, differing by only $1.71\%$, with 4639 and 4757 local passes, respectively. Nevertheless,
\[
\frac{\pi}{\sqrt{2}}
\rho_{\mathrm{AID},A}^{(K_A^\star)}
=0.0546,
\qquad
\frac{\pi}{\sqrt{2}}
\rho_{\mathrm{AID},B}^{(K_B^\star)}
=0.1328.
\]
Thus, B exhibits $2.44\times$ greater orientation-induced RMS variation. Its orientation gain is also $2.67\times$ larger, $0.3597$ versus $0.1349$. Hence, nearly equal placement-visible trajectory intensity does not imply comparable orientation sensitivity; placement changes the locally visible directional density $f_i$ and therefore the target--sensor harmonic coupling.

Figure~\ref{fig:aid_mechanism}(b) illustrates the shared-harmonic mechanism in Theorem~\ref{thm:shared}. At location B, the dominant shared harmonic is $m^\star=2$, reflecting the coupling between the prominent bidirectional vessel-motion structure and the corresponding sensor-response mode. Higher even harmonics capture finer directional structure. With $\eta=0.99$, \eqref{eq:Kstar} gives $K_B^\star=8$,
retaining $99.91\%$ of the resolved shared-mode strength.

As shown in Fig.~\ref{fig:aid_mechanism}(c), retaining only $K=1$ does not reproduce the orientation response, while $K=2$ captures its dominant two-lobed structure. Using $K_B^\star=8$ closely matches the directly evaluated response with RMSE $0.0039$. Thus, the same shared spectrum that determines orientation sensitivity also provides an accurate reduced-order representation of the orientation response.


\subsection{Historical-to-Future Robustness}

For each candidate, $\hat{\epsilon}_i$ in \eqref{eq:eps_cal} is calibrated from the five Sep.--Oct. windows, while the five Nov.--Dec. windows remain strictly held out, yielding $350\times5=1750$ candidate--window evaluations.

\begin{figure}[t]
\vspace*{0.05in}
\centering
\includegraphics[width=0.48\textwidth]{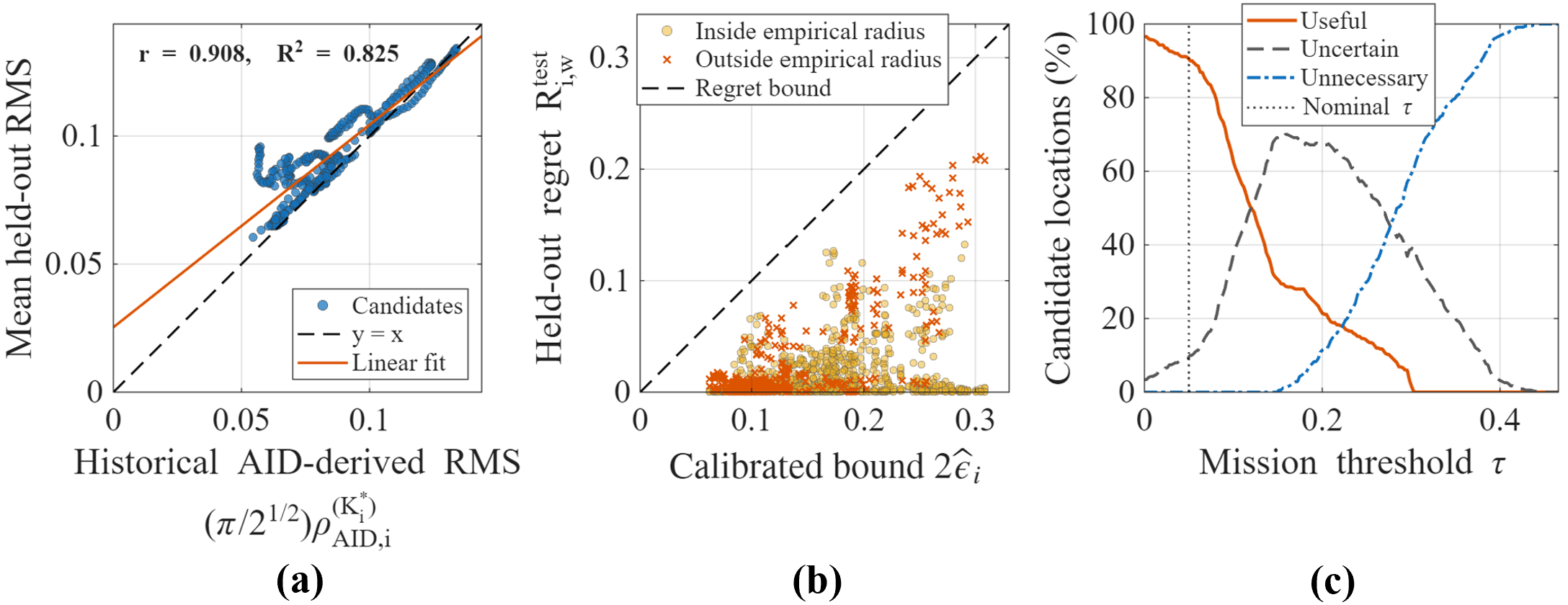}
\caption{Historical-to-future validation.
(a) Historical AID-derived RMS versus mean held-out RMS variation.
(b) Realized held-out orientation regret versus the calibrated bound
$2\hat{\epsilon}_i$; markers distinguish held-out future windows
inside and outside the empirical discrepancy radius.
(c) Fractions of candidate locations classified as
orientation-useful, uncertain, or orientation-unnecessary
versus mission threshold $\tau$; the vertical line indicates
the nominal threshold $\tau=0.05$.}
\label{fig:future}
\end{figure}

Figure~\ref{fig:future}(a) compares historical AID-derived RMS variation with the mean RMS variation across the held-out windows. Historical AID remains strongly predictive of future orientation sensitivity, with Pearson correlation $r=0.908$ and descriptive $R^2=0.825$.
This result is interpreted as temporal transfer of orientation sensitivity: locations that are highly orientation-sensitive historically tend to remain highly orientation-sensitive in the held-out period. It is not, by itself, evidence that AID outperforms target-only anisotropy when the sensor response is fixed across all locations.

The median empirical calibration radius is
$\operatorname{med}(\hat{\epsilon}_i)=0.0672$,
compared with a median held-out discrepancy of $0.0510$. The calibrated radius contains $77.1\%$ of the 1,750 held-out candidate--window discrepancies, and only $32.9\%$ of candidate locations remain inside the radius in all five held-out windows. These results confirm that $\hat{\epsilon}_i$ is an empirical calibration radius rather than a probabilistic confidence bound guaranteeing that every future distribution lies within the assumed uncertainty set.

Importantly, exceeding the empirical discrepancy radius does not necessarily imply poor orientation performance. The discrepancy $\mathcal D_i$ controls the entire orientation-response curve uniformly, whereas deployment regret depends only on the relative performance of the historical and future favorable orientations.

Figure~\ref{fig:future}(b) evaluates the orientation-regret result in Theorem~\ref{thm:robust}. For held-out window $w$, the realized regret from retaining the historical orientation is
\[
R_{i,w}^{\mathrm{test}}
=
d_{i,w}^{\mathrm{test}}
\!\left(\phi_{i,w}^{\mathrm{test},\star}\right)
-
d_{i,w}^{\mathrm{test}}
\!\left(\hat\phi_i^\star\right).
\]
The empirical inequality
\[
R_{i,w}^{\mathrm{test}}\leq 2\hat{\epsilon}_i
\]
holds in all 1,750 held-out candidate--window
evaluations. The median regret is $0.0031$,
and the median and 90th-percentile
regret-to-bound ratios are $0.024$ and $0.273$,
respectively.
The empirical bound remains conservative even
when the discrepancy radius is exceeded, because
regret depends on the selected orientations rather
than the maximum response-curve deviation.

Because $\hat{\epsilon}_i$ is estimated from calibration data rather than guaranteed to satisfy Assumption~\ref{ass:Dh} for every future distribution, in this example, the observed $100\%$ regret-bound consistency is an out-of-sample empirical result rather than a formal guarantee for arbitrary future periods.

At the nominal threshold $\tau=0.05$, 316 of the 350 candidate locations satisfy
$\hat G_i-\hat{\epsilon}_i>\tau,$
while 34 remain uncertain and no location satisfies $ 
\hat G_i+\hat{\epsilon}_i<\tau$.
Most importantly, all 316 locations satisfying the empirically calibrated sufficient condition retain normalized gain above $\tau$ when the historically selected orientation $\hat\phi_i^\star$ is applied in each of the five held-out future windows. Thus, the observed $316/316$ result directly evaluates the deployment-relevant statement in Corollary~\ref{cor:certificate}: the historical orientation itself, rather than only the unknown future-optimal orientation, remains useful in the held-out data. Because the candidate locations and temporal windows are not independent realizations and $\hat{\epsilon}_i$ is empirically estimated, this observation should not be interpreted as a $100\%$ probabilistic guarantee for arbitrary future deployments.

Figure~\ref{fig:future}(c) examines sensitivity to the mission-selected threshold $\tau$. As $\tau$ increases, the fraction of orientation-useful locations decreases, the fraction for which orientation selection is unnecessary increases, and the uncertain class dominates an intermediate range. The nominal value $\tau=0.05$ lies in the low-threshold regime, explaining why no orientation-unnecessary cases occur at the nominal operating point. The extended threshold sweep is diagnostic and is computed only from historical and calibration quantities; held-out data are not used to select $\tau$.

We also assess calibration-window sensitivity by dividing
the Sep.--Oct. calibration period into 3, 5, and 10 windows
while retaining the same five held-out test windows.
The median calibrated radius increases from 0.0511 to
0.0672 and 0.0998, with held-out discrepancy coverage
of 52.9\%, 77.1\%, and 98.6\%, respectively.
The corresponding numbers of orientation-useful locations
are 321, 316, and 263, illustrating a trade-off between
empirical uncertainty coverage and decision conservatism.
All classified useful locations retain gain above
$\tau=0.05$ across all five held-out windows. 
Because the calibration period is fixed, varying
the window count also changes window duration;
this experiment therefore assesses temporal-resolution
sensitivity rather than calibration sample size alone.

\subsection{Baselines and Sensor-Response Ablation}

We first assess whether conditioning distribution shift on the sensing task reduces conservatism. Using the same historical and calibration directional distributions, the median calibrated radius is $0.0672$ for the proposed sensor-induced discrepancy $\mathcal D_i$, compared with $0.2360$ for total-variation distance. Both yield $100\%$ observed held-out regret-bound consistency; however, $\mathcal D_i$ yields 316 orientation-useful decisions and leaves 34 cases uncertain, whereas total variation yields only 89 orientation-useful decisions and leaves 261 uncertain. Thus, incorporating the sensor response primarily reduces conservatism rather than merely altering observed held-out coverage.

We next examine whether AID provides information beyond target anisotropy alone. For the fixed $60^\circ$ angular-acceptance response, target-only spectral anisotropy performs similarly to AID, with $R^2=0.835$ versus $0.825$. This is expected because the same sensor spectrum $H_{i,m}$ applies to every candidate location; consequently, the fixed-response comparison alone does not establish an advantage of AID over target-only anisotropy.

The distinction predicted by Theorem~\ref{thm:shared} becomes apparent when the sensor response itself changes. We repeat the analysis using the same target trajectories and candidate locations while varying the total angular-acceptance width over $30^\circ, 60^\circ, 90^\circ, 120^\circ$. Across this sweep, the median AID-derived RMS increases from $0.0500$ to $0.1102$, the median retained order $K_i^\star$ decreases from 12 to 4, and the number of orientation-useful decisions increases from 249 to 332.

Pooling 350 candidate locations across four
sensor-response widths, linear regressions against
mean held-out orientation-response RMS yield
$R^2=0.920$ for historical AID-derived RMS and $R^2=0.239$ for historical target-only spectral
anisotropy, measured by the unweighted spectral
magnitude
$\left(\sum_{m=1}^{K_{\max}}F_{i,m}^{2}\right)^{1/2}$
with $K_{\max}=18$. These descriptive results demonstrate
stronger historical-to-future association when
sensor-response-dependent harmonic coupling is
included, without implying independent observations
across sensor widths.

These results directly evaluate sensitivity to the assumed sensing characteristics within the separable model. Changing $h_i$ alters both the magnitude and spectral complexity of orientation sensitivity even when the target trajectories are unchanged, supporting the shared dependence $F_{i,m}H_{i,m}$ in Theorem~\ref{thm:shared}. The width sweep does not establish generality to sensors whose spatial visibility footprint itself changes with orientation, which remain outside the model scope described in Section~\ref{sec:model}.



\section{Conclusion}
\label{sec:conclusion}

This letter characterizes when sensor orientation matters after placement is fixed. Orientation sensitivity arises from harmonic modes shared by the locally visible target-direction distribution and sensor response, with AID providing an exact RMS measure of its retained-harmonic variation. A sensor-induced directional discrepancy provides
conditional bounds on future orientation regret and
retained gain, while data-calibrated radii support
empirical orientation-selection decisions rather than
unconditional future guarantees. AIS experiments support both the shared-harmonic mechanism and historical-to-future orientation transfer. 

\balance
\bibliographystyle{IEEEtran}
\bibliography{main}

\end{document}